\documentclass[11pt]{article}

\usepackage{color,latexsym,amsmath,amssymb,path,url,enumerate}
\usepackage[colorlinks=false,bookmarks=false,hyperfootnotes=false]{hyperref}
\newcommand{\ignore}[1]{}

\newtheorem{theorem}{Theorem}[section]
\newtheorem{lemma}[theorem]{Lemma}
\newtheorem{corollary}[theorem]{Corollary}
\newtheorem{observation}[theorem]{Observation}

\newcommand{\Proof}[1]
        {
        \noindent
        \emph{Proof #1.}~
        }
\newsavebox{\smallProofsym}                     % smallproofsym.tex
\savebox{\smallProofsym}                        %
        {%                                      %
        \begin{picture}(7,7)                    %
        \put(0,0){\framebox(6,6){}}             %
        \put(0,2){\framebox(4,4){}}             %
        \end{picture}                           %
        }                                       %
\newcommand{\smalleop}[1]
        {
        \mbox{} \hfill #1~~\usebox{\smallProofsym}\!\!\!\!\!\!\
        }
\newenvironment{theProof}[1]
        {
        \Proof{#1}}{\smalleop{}
        \medskip

        }
\usepackage{graphicx,psfrag}
\usepackage{dsfont}

\newcommand{\tri}[1]{{\mathsf{tr}}(#1)}
\newcommand{\pt}[1]{{\mathsf{pt}}(#1)}
\newcommand{\ppt}[1]{{\mathsf{ppt}}(#1)}
\newcommand{\ptw}[2]{{\mathsf{pt}_{#1}}(#2)}

\begin{document}

\title{On Numbers of Pseudo-Triangulations\thanks{Work on this paper by the third author was partially supported by Grant 338/09 from
the Israel Science Fund and by the Israeli Centers of Research Excellence (I-CORE) program (Center  No. 4/11).}}

\author{
Moria Ben-Ner\thanks{%
School of Computer Science, Tel Aviv University, Tel Aviv 69978, Israel\@.
Supported by Grant 338/09 from
the Israel Science Fund.
Email: \texttt{moriaben@tau.ac.il} }
\and
Andr\'e Schulz\thanks{%
Institut f\"ur Mathematische Logik und Grundlagenforschung, Universit\"at M\"unster,
Email: \texttt{andre.schulz@uni-muenster.de} }
\and
Adam Sheffer\thanks{%
School of Computer Science, Tel Aviv University, Tel Aviv 69978, Israel\@.
Supported by Grant 338/09 from
the Israel Science Fund.
Email: \texttt{sheffera@tau.ac.il} }
}

\maketitle
\pagenumbering{arabic}

\begin{abstract}
We study the maximum numbers of pseudo-triangulations and pointed pseudo-triangulations that can be embedded over a specific set of points in the plane or contained in a specific triangulation.

We derive the bounds $O(5.45^N)$ and $\Omega (2.41^N)$ for the maximum number of pointed pseudo-triangulations that can be contained in a specific triangulation over a set of $N$ points. For the number of all pseudo-triangulations contained in a  triangulation we derive the bounds $O^*(6.54^N)$ and $\Omega (3.30^N)$.
We also prove that $O^*(89.1^N)$ pointed pseudo-triangulations can be embedded over any specific set of $N$ points in the plane, and at most $120^N$ general pseudo-triangulations.
\end{abstract}

%%%%%%%%%%%%%%%%%%%%%%%%%%%%%%%%%%%%%%%%%%%%%%%%%%%%%%%%%%% SECTION
\section{Introduction}

\begin{figure}[t]
\begin{center}
\includegraphics[width=0.45\textwidth]{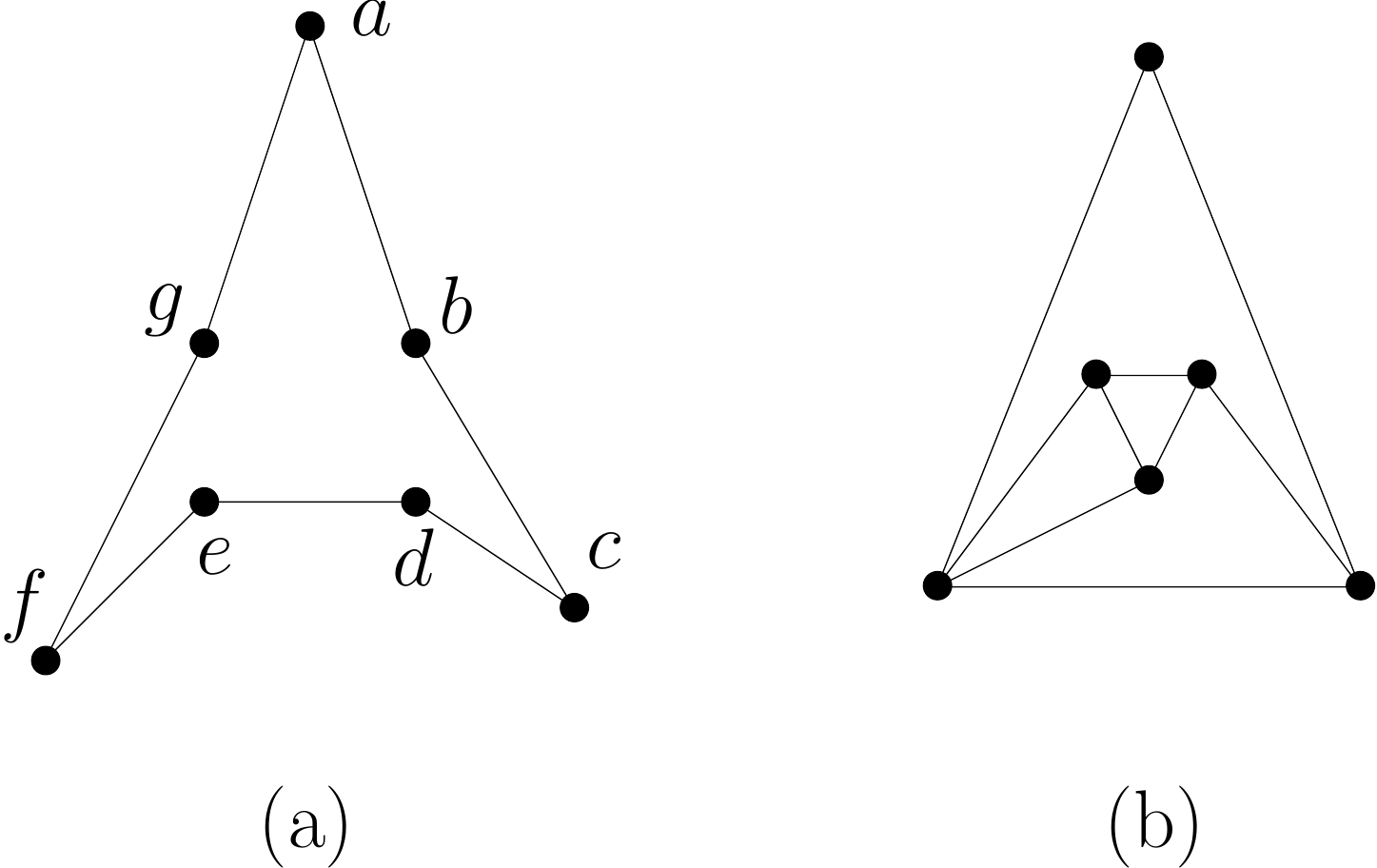}
\end{center}
\vspace{-6mm}

\caption{\small \sf (a) A pseudo-triangle.
(b) A pseudo-triangulation tiles the convex hull of the point set into pseudo-triangles.
}
\label{fi:counter}
\end{figure}

%Introduce PT
A geometric graph is a graph whose vertices are associated with points in the plane and whose edges are represented as straight line-segments.
A \emph{pseudo-triangle} is a simple polygon with exactly three convex vertices, called \emph{corners} (e.g., see Figure~\ref{fi:counter}(a)).
A \emph{pseudo-triangulation} is a crossing-free connected geometric graph that contains the edges of the convex hull of its vertices, and whose interior faces are all pseudo-triangles (e.g., see Figure~\ref{fi:counter}(b)). A pseudo-triangulation is \emph{pointed} if every vertex is incident to an angle larger than $\pi$; for example, the pseudo-triangulation in Figure~\ref{fi:counter}(b) is pointed. Pseudo-triangulations have many interesting properties and various applications as geometric data structures in areas such as motion planning, polygon unfolding, and ray shooting. For a comprehensive list of applications, see the survey of Rote, Santos, and Streinu~\cite{RSS06}. Interestingly, pointed pseudo-triangulations are exactly the planar generically minimal rigid graphs in the plane~\cite{ROR+03,S05}. By Laman's theorem, generically minimal rigid graphs in the plane can be described by a combinatorial condition~\cite{L70}. Therefore, despite of their geometric structure, the ``graphs'' of pointed pseudo-triangulations have a purely combinatorially description.

% Scope of the paper
In this paper we address the question of how many (pointed and non-pointed) pseudo-triangulations can be embedded on a specific point set, or are contained in a specific geometric triangulation. We consider the graphs to be labeled (that is, two geometric graphs are considered different if they differ in at least one edge, even when they are isomorphic). We also assume that the point sets are in general position, in the sense that no three points are collinear. For the purpose of bounding maximum numbers of crossing-free geometric graphs this involves no loss of generality, since such numbers can only grow when a degenerate point set is slightly perturbed into general position.
For simplicity, we refer to crossing-free geometric graphs as \emph{plane graphs}.

%Counting plane graphs
The problem of bounding the maximum number of plane graphs on a set of $N$ points has a 30 year long history. The first major result, derived by Ajtai et al.~\cite{ACNS82}, was an upper bound of $10^{13N}$ for the number of plane graphs that can be embedded over a specific set of $N$ points. Since then, a constant flow of improvements culminated in the recent upper bound of $O^*(187.53^N)$ by Sharir and Sheffer~\cite{SS12}.\footnote{In the notations $O^*()$, $\Theta^*()$, and $\Omega^*()$, we neglect polynomial factors.} In Table~\ref{tab:1}, we list the current bounds for several common variants of this problem. A thorough survey of these and other variants can be found in Aichholzer~et~al.~\cite{AHHHKV07}.\footnote{We  maintain up-to-date results in \url{http://www.cs.tau.ac.il/~sheffera/counting/PlaneGraphs.html} (version of October 2012).} Most of the upper bounds were obtained by estimating the maximum (or expected) number of graphs (of the respective type) that can be contained in a single triangulation. One reason for this is that the current upper bound for the number of triangulations is relatively small. Moreover, the property that every plane graph is contained in at least one triangulation makes it easier to obtain, for the various variants, bounds that rely on the maximum number of triangulations. As a consequence, upper bounds for maximum numbers of graphs contained in a single triangulation play an important role in this field.
%Maybe add: Motivation on counting: bounds for enumeration
% Table: triangulations, plane graphs, spanning trees cycles, ham cycles, matchings, perfect matchings, pt, ppt with references
%
\begin{table}[h]
\begin{center}
\begin{tabular}{|c|c|c|}
\hline
 {\textbf {Graph class}} & {\textbf {Lower bound}}
& {\textbf {Upper bound}} \\
\hline\hline
 Plane graphs &
$\Omega(41.18^N)$~\cite{AHHHKV07,GNT00} & $O^*(187.53^N)$~\cite{SS12}\\
\hline
 Cycle-free graphs & $\Omega(12.26^N)$~\cite{DSST11} & $O(160.55^N)$~\cite{HSSTW11,SS10}\\
\hline
  Perfect matchings & $\Omega^*(3^N)$~\cite{GNT00} & $O(10.07^N)$~\cite{SW06} \\
\hline
  Spanning trees & $\Omega(12.00^N)$~\cite{DSST11} & $O(141.07^N)$~\cite{HSSTW11,SS10} \\
\hline	
 Spanning cycles & $\Omega(4.64^N)$~\cite{GNT00} & $O(54.55^N)$~\cite{SSW12} \\
\hline
 Triangulations & $\Omega(8.65^N)$~\cite{DSST11} & $O(30^N)$~\cite{SS10} \\
\hline
\end{tabular}
\caption{\small Current bounds for maximum number of graphs that can be embedded over a point set.\label{tab:1}}
\end{center}
\vspace{-\baselineskip}
\end{table}

The motivation for deriving bounds for the multiplicity of plane graphs stems from the need to understand the complexity of non-crossings graphs. For example, how many bits are needed to represent a triangulation, or what is the running time of an algorithm that extensively searches through all plane graphs of a certain type. Notice that in many geometric optimization problems the solution has to be necessarily a plane graph. In particular, the minimum Euclidean spanning tree and the minimum Euclidean TSP tour are both necessarily crossing-free, and it is not hard to see that the minimum Euclidean minimal rigid graph has to be crossing-free as well.

In the spirit of the bounds displayed in Table~\ref{tab:1}, recent works cite the following upper bounds for maximum numbers of (pointed) pseudo-triangulations~\cite{AHHHKV07,Vog07} (all attributed to Randell~et~al.~\cite{RRSS01}):\begin{enumerate}[(i)]\vspace{-2.5mm}

\item \label{it:wrongPTinTR}Every triangulation embedded over a set of $N$ points in the plane contains $O(3^N)$ pseudo-triangulations. This result, together with the following one, are stated in Table 2 of \cite{AHHHKV07}.\vspace{-2.5mm}

\item \label{it:wrongPPTinTR}Every triangulation embedded over a set of $N$ points in the plane contains $O(3^N)$ \emph{pointed} pseudo-triangulations. \vspace{-2.5mm}

\item \label{it:wrongPTvsTR}The maximum number of pseudo-triangulations that can be embedded over a specific set $S$ of $N$ points is at most
the number of triangulations that can be embedded over $S$ times $O(3^N)$. This result, together with the following one, are stated in Table 3 of \cite{AHHHKV07}\footnote{More specifically, it is stated in the table that the maximum number of pseudo-triangulations that can be embedded over a set of $N$ planar points is $O(129^N)$. The basis 129 is obtained by multiplying 3 with 43, which, at the time \cite{AHHHKV07} was published, was the best known upper bound on the maximum number of triangulations that can be embedded over a set of $N$ planar points.} (this is an immediate corollary of \eqref{it:wrongPTinTR}). \vspace{-2.5mm}

\item \label{it:oldPPTvsTR}The maximum number of \emph{pointed} pseudo-triangulations that can be embedded over a specific set $S$ of $N$ points is at most
the number of triangulations that can be embedded over $S$ times $O(3^N)$ (this is an immediate corollary of \eqref{it:wrongPPTinTR}).
\end{enumerate}

\noindent However, while \eqref{it:oldPPTvsTR} is indeed proved by Randell~et~al.~\cite[Theorem 8]{RRSS01}, the other three bounds are not proved in it, and do not seem to be directly implied by it. In fact, bound~(\ref{it:wrongPTinTR}) does not hold, as proved in Section~\ref{sec:past}.\footnote{The problematic bounds are a rather small part of \cite{AHHHKV07}, and the many other result in that paper are not affected by it.}
%In Section \ref{sec:past}, we discuss the problems with the above bounds, disprove \eqref{it:wrongPTinTR}, and present %some first new bounds.
Specifically, we derive a lower bound of $\Omega (3.30^N)$ on the maximum number of pseudo-triangulations contained in a single triangulation, and a  lower bound of $\Omega (2.41^N)$ on the maximum number of \emph{pointed} pseudo-triangulations contained in a single triangulation.

In Section~\ref{sec:inTR} we derive a bound of $O(5.45^N)$ for  the maximum number of pseudo-triangulations contained in a single triangulation, and a bound of $O^*(6.54^N)$ for the the maximum number of pointed pseudo-triangulations. We obtain these bounds by relying on an approach completely different from the one presented by Randell~et~al.~\cite{RRSS01}.

In Section~\ref{sec:conj}, we present a new observation concerning numbers of pseudo-triangulations, and use it to show that bound \eqref{it:oldPPTvsTR} is not asymptotically tight. Although we only slightly decrease the bound from $3^N$ down to $2.97^N$, this is the first dent made to this bound in a decade, and it seems likely that our approach has potential to induce further progress.

The observations in Sections~\ref{sec:past}~and~\ref{sec:conj} also imply that at most $120^N$ general pseudo-triangulations can be embedded over any specific set of $N$ points in the plane, and $O^*(89{.}1^N)$ pointed pseudo-triangulations. The current best \emph{lower bounds} are $\Omega(12^N)$ for pointed pseudo-triangulations, and $\Omega(20^N)$ for general pseudo-triangulations~\cite{AHHHKV07}. Both lower bounds were obtained by counting pseudo-triangulations of a point set which consists of two opposing concave chains (the \emph{double chain} configuration). The bounds for the number of pseudo-triangulations of the double chain are known to be asymptotically tight (ignoring polynomial factors). \vspace{2mm}

\noindent{\bf Notation.}
For a set $S$ of points in the plane, we denote by $\tri{S}$ the number of distinct triangulations that can be embedded over $S$.
Moreover, we let $\tri{N}=\max_{|S|=N}\tri{S}$.
Similarly, we respectively denote by $\pt{S}$ and $\ppt{S}$ the number of pseudo-triangulations and pointed pseudo-triangulations that can be embedded over $S$, and set $\pt{N}=\max_{|S|=N}\pt{S}$, and $\ppt{N}=\max_{|S|=N}\ppt{S}$.

When considering a point set $S$, we denote as $S' \subset S$ the set of all points that are not contained in the boundary of the convex hull of $S$; we denote such points as \emph{interior points}.
Moreover, we let $N = |S|$, $n =|S'|$, and $h=N-n$. For every $W \subseteq S'$, we denote as $\ptw{W}{S}$ the number of pseudo-triangulations over the point set $S$
that have $W$ as the set of their pointed interior vertices. For example, $\ptw{\emptyset}{S} = \tri{S}$, $\ptw{S'}{S} = \ppt{S}$, and $\sum_W \ptw{W}{S} = \pt{S}$.

% Maybe put this as a footnote when first used?
We will also use the following notation --- given two plane graphs $G$ and $H$ over the same point set $S$, if every edge of $G$ is also an edge of $H$, we write $G \subseteq H$.

\section{Known results and first bounds} \label{sec:past}
\begin{figure}[htb]
\begin{center}
\begin{tabular}{cccc}
  \includegraphics[width=.21\columnwidth,page=1]{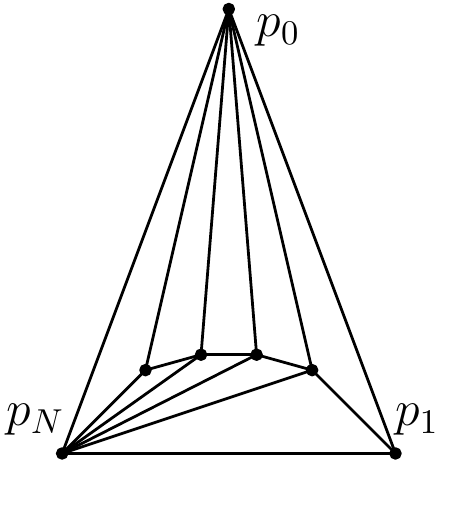} &
  \includegraphics[width=.21\columnwidth,page=2]{lowerbound} &
  \includegraphics[width=.21\columnwidth,page=3]{lowerbound} &
  \includegraphics[width=.21\columnwidth,page=4]{lowerbound}
  \\
\small{$T_6$} &  \small{case A} &  \small{case B} &  \small{case C}
\end{tabular}
\caption{The triangulation $T_6$ and cases A,B,C with a right pseudo-triangle of width 2, as discussed in the proof of Theorem~\ref{thm:lowerbound}. Case C is relevant only when counting non-pointed pseudo-traingulations.}
  \label{fi:lowerbound}
\end{center}
\end{figure}

Let us first present a counterexample for bound \eqref{it:wrongPTinTR}.
Consider an $N+1$ elementary point set that forms a convex chain with one additional point far above the chain, such that the convex hull of the set is triangular. We denote the topmost point as $p_0$, and the points on the chain as $p_1,\cdots,p_N$, from right to left.
Let $T_N$ be the triangulation of this set as depicted in the leftmost part of Figure~\ref{fi:lowerbound}, and let $P_N$ denote the number of pointed pseudo-triangulations that are contained in $T_N$. Similarly, denote by $Q_N$ the number of  pseudo-triangulations that are contained in $T_N$.
%
%Next, we analyze how many pseudo-triangulations are contained in each of the above 6-point configurations (i.e., non-convex quadrilaterals with two interior vertices of degree 3).
%For each of the two interior vertices of degree 3, we can either remove one of the three edges adjacent to it, or leave it as a vertex of degree 3.
%We thus obtain $4\cdot4=16$ pseudo-triangulations.
%Each of those remains a pseudo-triangulation after the removal of the diagonal of the quadrilateral (the edge $ab$ in Figure \ref{fi:counter}(d)),
%implying that there are 32 pseudo-triangulations contained in each 6-point configuration. It can be easily check that when the set contains $N$ points,
%there are $N/3-1$ such configurations.
%Therefore, the number of pseudo-triangulations that are contained in the triangulation is $\Omega (32^{N/3}) = \Omega (3.17^N)$.

%
% Add Figures here
\begin{theorem}\label{thm:lowerbound}
The triangulation $T_N$ contains $\Theta\left((\sqrt{2}+1)^N\right)$ pointed pseudo-triangulations and $\Theta\left(\left(\frac{\sqrt{13}+3}{2}\right)^N\right)$ pseudo-triangulations.
\end{theorem}
\begin{theProof}{\!\!\!}
For every pseudo-triangulation in $T_N$ we call the pseudo-triangle that contains the edge $p_0p_1$ the \emph{right pseudo-triangle}. When $p_0p_{k+1}$ is part of the right pseudo-triangle, we say that the right pseudo-triangle has \emph{width} $k$. The width of the right pseudo-triangle can range between $1$ and $N-1$. We count the number of pointed pseudo-triangulations that are contained in $T_N$ and have a right pseudo-triangle of width $k$. Since $p_{k+1}$ has to be pointed, the vertices of the right pseudo-triangle (in order) are either $p_0,p_1,\cdots, p_k,p_N,p_{k+1}$ (case A in Figure~\ref{fi:lowerbound}), or  $p_0,p_1,\cdots, p_k,p_{k+1}$ (case B in Figure~\ref{fi:lowerbound}). In case A we can pick every pointed pseudo-triangulation out of $T_N$ restricted to $p_{k+1},p_{k+2},\ldots,p_N,p_0$ to construct a pointed pseudo-triangulation for $T_N$. This set induces a triangulation of the form $T_{N-k}$ (e.g., see the shaded region in Figure~\ref{fi:lowerbound}). A similar situation occurs in case B, where we can pick any pointed pseudo-triangulation contained in $T_{N-k-1}$ to define a pseudo-triangulation for $T_N$. By also noting that there is a single pointed pseudo-triangulation when $k=N-1$, we obtain
\begin{align*}
P_N &=  1 + \sum_{k=1}^{N-2} \left( P_{N-k} + P_{N-k-1} \right) \\
      &=  1 + P_1 + P_{N-1} + 2\sum_{k=2}^{N-2} \ P_{N-k}  \\
        &=  P_{N-1} + P_{N-2} + \left(1 + P_1 + P_{N-2} +2\sum_{k=3}^{N-2} \ P_{N-k} \right) \\
      &=  2P_{N-1}+ P_{N-2}.
\end{align*}
Solving the recurrence with $P_1=1$ and $P_2=1$ gives
\begin{align*}
P_N:= \frac{\sqrt{2}-1}{2} \left(\left(1+\sqrt{2}\right)^N-\left(3+2\sqrt{2}\right)\left(1-\sqrt{2}\right)^N\right)=\Theta\left(\left(\sqrt{2}+1\right)^N\right).
\end{align*}

To count the number of all pseudo-triangulations we slightly modify the counting scheme. We sum up again all pseudo-triangulations whose width of the right pseudo-triangle is $k$, while $k$ ranges from $1$ to $N-1$. As in the previous analysis, we have cases A and B, but this time $p_{k+1}$ may also be non-pointed. This results in a new case, called C, which is depicted in Figure~\ref{fi:lowerbound}. In this case we may pick any pseudo-triangulation from $T_{N-k}$ to obtain a pseudo-triangulation for $T_N$. Thus, we have
\begin{align*}
Q_N &=  1+ \sum_{k=1}^{N-2} \left(2 Q_{N-k} + Q_{N-k-1} \right) \\
      &=  1+ Q_1 + 2Q_{N-1} + 3\sum_{k=2}^{N-2} \ Q_{N-k} \\
      &= 2 Q_{N-1} + Q_{N-2} + (1 + Q_1 + 2 Q_{N-2} +3\sum_{k=3}^{N-2} \ Q_{N-k}) \\
      &=  3Q_{N-1}+ Q_{N-2}.
\end{align*}
Solving the  recurrence with $Q_1=1$ and $Q_2=1$ gives
\begin{align*}
Q_N &= \frac{\left(\left(13-\sqrt{13}\right)\left(3+\sqrt{13}\right)^N - \left(91+25\sqrt{13}\right)\left(3-\sqrt{13}\right)^N\right) }{13 \left(3+\sqrt{13}\right)2^N} \\[2mm]
&= \Theta\left((3+\sqrt{13})/2)^N\right).
\end{align*} %\vspace{-10mm}
\end{theProof}

\begin{corollary}
$\ppt{N} =\Omega(2.41^N)$, $\pt{N} =\Omega(3.30^N)$.
\end{corollary}

In the remainder of this section we show why bound \eqref{it:wrongPTinTR} is not implied by~\cite[Theorem 8]{RRSS01}. Let us first repeat the theorem.
\begin{theorem} \label{le:old} {\bf \cite{RRSS01}}
Given a set $S$ of $N$ points in the plane, a subset $W \subseteq S'$, and a point $v \in W$, then $\ptw{W}{S} \le 3\cdot \ptw{W\setminus \{v\}}{S}$.
\end{theorem}
\begin{theProof}{sketch}
The proof is obtained by combining two simple observations:

1. For every pseudo-triangulation with $W$ as the set of its pointed interior vertices, we can insert a (unique) single edge to obtain a pseudo-triangulation
with $W \setminus \{v\}$ as the set of its pointed interior vertices.

2. For every pseudo-triangulation with $W \setminus \{v\}$ as the set of its pointed interior vertices, there are at most three edges whose removal will
 form a pseudo-triangulation with $W$ as the set of its pointed interior vertices (the removal of a single edge. Not of the entire set).
\end{theProof}

Bound \eqref{it:oldPPTvsTR} is a simple corollary of Theorem \ref{le:old}, since
\begin{align*}
\ppt{S} = \ptw{S'}{S} \le 3 \cdot \ptw{S' \setminus \{v_1\}}{S} \le 3^2 \cdot \ptw{S' \setminus \{v_1,v_2\}}{S} \le \ldots &\le 3^n \cdot \ptw{\emptyset}{S} \\
&< 3^N \cdot \tri{S},
\end{align*}
where $S'=\{v_1, v_2, ..., v_n\}$. However, bounds \eqref{it:wrongPTinTR} and \eqref{it:wrongPPTinTR} cannot be obtained in a similar manner,
since the first observation of the proof is no longer valid when considering a specific triangulation.
For example, consider the pseudo-triangle depicted
in Figure \ref{fi:counter}(a), and assume that we wish to add an edge that will make $e$ non-pointed. This would require the insertion of the edge $ae$
(inserting a different  edge will either leave $e$ pointed or will not result in a valid pseudo-triangulation), which is possible
when dealing with bound \eqref{it:oldPPTvsTR}. However, when considering pseudo-triangulations that are contained in a specific triangulation $T$, this might not
work since $T$ might not contain the edge $ae$ (for example, $T$ might contain the edges $gb,gd,$ and $ge$ instead).

Finally, consider bound \eqref{it:wrongPTvsTR}, which claims that for every set $S$ of $N$ points in the plane, $\pt{S} =  O^*\left(3^N \cdot \tri{S}\right)$. Notice that $\pt{S} = \sum_W \ptw{W}{S}$. While Theorem \ref{le:old} implies that $\ptw{W}{S} \le 3^{|W|} \cdot \tri{S}$ holds for every $W$, it does not imply bound \eqref{it:wrongPTvsTR}, since there is an exponential number of terms in the above sum. For a very simple-minded bound, we note that there are fewer than $2^N$ terms in the sum, so we get $\pt{S} < 6^N \cdot \tri{S}$. This can be easily improved, as follows.
\begin{theorem} \label{th:PTvsTR}
For every set $S$ of $N$ points in the plane, $\pt{S} < 4^N \cdot \tri{S}$.
\end{theorem}
\begin{theProof}{\!\!}
Given a specific set $W$ that contains exactly $k$ interior vertices, Theorem \ref{le:old} implies $\ptw{W}{S} \le 3^k \cdot \tri{S}$.
Moreover, there are $\binom{n}{k} < \binom{N}{k}$ such sets.
By combining the above with the binomial theorem, we obtain
\[ \pt{S} = \sum_W \ptw{W}{S} < \sum_{k=0}^{N} \binom{N}{k} \cdot 3^k \cdot \tri{S} = \tri{S} \cdot \sum_{k=0}^{N} \binom{N}{k} \cdot 3^k  = 4^N \cdot \tri{S}. \] \vspace{-4mm}

\end{theProof}
By combining Theorem \ref{th:PTvsTR} with the bound $\tri{N} < 30^N$ from \cite{SS10}, we obtain
\begin{corollary}
$\pt{N} < 120^N$.
\end{corollary}
\section{The number of pseudo-triangulations in a triangulation} \label{sec:inTR}

In this section we use a new technique to obtain upper bounds for the maximum numbers of pseudo-triangulations and pointed pseudo-triangulations that
are contained in a specific triangulation. Our technique relies on the following lemma. Interestingly, after completing this section, we noticed that the lemma was already proved by Rote~et~al.~\cite[Theorem 1]{RWWX03} (where it is used for different purposes).
We present our proof here, for completeness, and since it is completely different from the one in~\cite{RWWX03}.
\begin{lemma} {\bf \cite{RWWX03}} \label{le:bijection}
Given a triangulation $T$ and a pseudo-triangulation $T' \subseteq T$, let $D$ be the set of edges that are in $T$ but not in $T'$,
and let $P$ be the set of pointed interior vertices of $T'$. Then there is a bijection between $D$ and $P$, such that every vertex of $P$ is adjacent to the edge in $D$  that corresponds to it.
\end{lemma}
\begin{theProof}{\!\!}
Consider a pseudo-triangle $\Delta$ of $T'$ with $k$ pointed vertices whose reflex angles are inside of $\Delta$.
It can be easily shown that there are $k$ diagonals of $\Delta$ that are in $T$ but not in $T'$ (in fact, exactly $k$ diagonals are required to triangulate any simple
polygon with $k+3$ vertices).
Since every pointed interior vertex has a reflex angle in a single pseudo-triangle, $D$ and $P$ are of the same size.
We prove the following stronger property: for each pseudo-triangle $\Delta$ of $T'$, there is a bijection between the diagonals of $\Delta$ in $T$ and the vertices of $\Delta$ that have a reflex angle in $T'$.
\begin{figure}[h]
\centering
\includegraphics[width=0.9\textwidth]{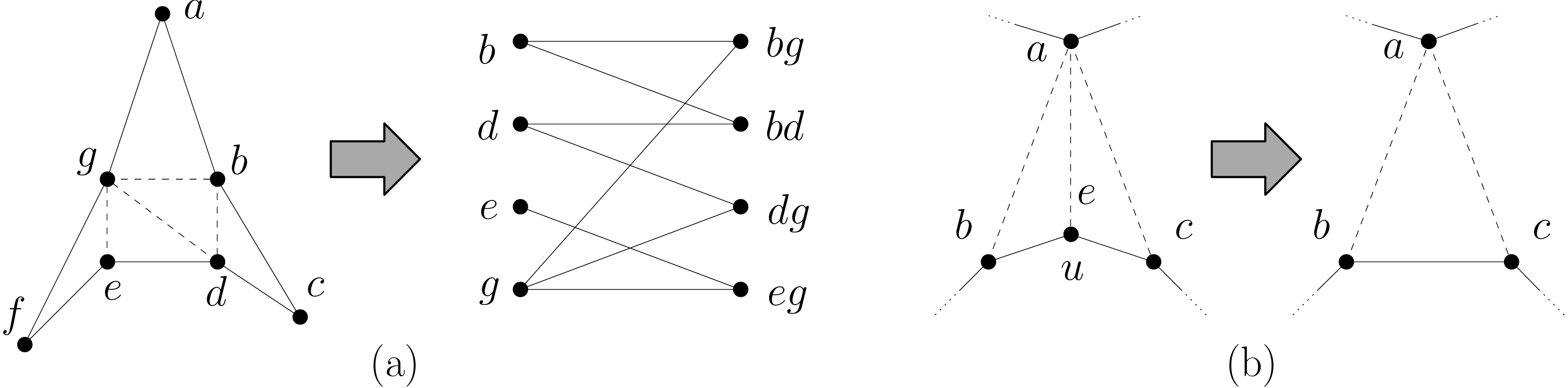}
\caption{\small \sf (a) A pseudo-triangle with four pointed vertices and four diagonals, and its corresponding bipartite graph.
(b) Removing a vertex $u$ adjacent to a single diagonal $e$.}
\label{fi:perfect}
\end{figure}

Consider a pseudo-triangle $\Delta$ of $T'$. We build a bipartite graph $G=(V_1,V_2,E)$, where $V_1$ contains a vertex for every pointed vertex of $\Delta$,
$V_2$ contains a vertex for every diagonal of $\Delta$, and there exists an edge $(v,u) \in E$ such that $v\in V_1$ and $u \in V_2$ if the vertex
that corresponds to $v$ is incident to the edge that corresponds to $u$. An example of such a graph is depicted in Figure \ref{fi:perfect}(a).
Notice that proving the existence of a bijection in $\Delta$ is equivalent to proving the existence of a perfect matching in $G$.

Hall's Theorem \cite[Theorem 1]{Hall35} states that $G$ contains a perfect matching if (and only if) for each subset $S_1 \subseteq V_1$, it holds that $|S_1| \le |S_2|$,
where $S_2 \subseteq V_2$ is the subset of vertices in $S_2$ that are adjacent to at least one vertex of $S_1$. Since every vertex in $V_2$ has a degree of at most 2 (a degree of 1 is obtained when one of the endpoints of the corresponding edge is connected to a corner), the pigeonhole principle implies that the condition holds for every subset $S_1$ with no vertex of degree 1.

To deal with vertices of degree 1, we prove the existence of a bijection in a pseudo-triangle $\Delta$ by induction on the number of the diagonals inside of it. If $\Delta$ contains a single diagonal (and thus a single pointed vertex), the claim obviously holds. Next, assume that the claim holds for every pseudo-triangle with $k-1$ diagonals, and consider a pseudo-triangle $\Delta$ with $k$ diagonals.
If $\Delta$ contains no pointed vertices with only a single diagonal adjacent to them, the claim is implied by Hall's theorem. Otherwise, we take such a pointed vertex $u$ and say that it corresponds to the only diagonal $e$ adjacent to it. We then remove $u$ and $e$, and obtain a valid pseudo-triangle by connecting the two neighbors of $u$ along the boundary of $\Delta$, as depicted in Figure \ref{fi:perfect}(b). Since we now have a pseudo-triangle with $k-1$ diagonals, the claim holds by the induction hypothesis.
\end{theProof}
Next, we use Lemma \ref{le:bijection} to obtain upper bounds for the maximum numbers of pseudo-triangulations and pointed pseudo-triangulations that
are contained in a specific triangulation.
\begin{theorem} \label{th:PPTinTR}
Any triangulation $T$ embedded over a set $S$ of $N$ points in the plane contains $O(5.45^N)$ pointed pseudo-triangulations.
\end{theorem}
\begin{theProof}{\!\!}
 By Lemma \ref{le:bijection}, every pointed pseudo-triangulation that is contained in $T$ can be obtained by iterating over the $n$ interior vertices of $T$ and for each vertex removing a single edge adjacent to it. Let us denote the degrees of the interior vertices as $d_1, \ldots, d_n$, and set $d_{n+1}=d_{n+2} = \ldots = d_N = 1$. The number of pointed pseudo-triangulations that are contained in $T$ is at most $\prod_{i=1}^{N}d_i$. Since a triangulation has fewer than $n+2N$ edges, $\sum_{i=1}^{N}d_i < 6N$. Combining this with the arithmetic-geometric mean inequality implies that the number of pointed pseudo-triangulations that are contained in $T$ is at most
\[\prod_{i=1}^{N}d_i \le \left(\frac{\sum_{i=1}^{N}d_i}{N}\right)^N < 6^N.\]
One inefficiency of the above method is that it counts many graphs where some edge was chosen at both of its endpoints.
Such graphs cannot be pointed pseudo-triangulations, since they contain too many edges.
To deal with this inefficiency, we apply an LP-based technique.
While this technique was used by Buchin and Schulz~\cite{BS10} to bound the maximum number of spanning trees in a single triangulation (embedded over a set of $N$ points in the plane), it can also be applied for our purposes.
In Section~\ref{sec:Orient}, we prove that there are $O(5.45^N)$ distinct ways for choosing a corresponding edge for every vertex, such that no edge is chosen more than once. This immediately implies the assertion of the theorem.

\ignore{To deal with this inefficiency, we notice that there exists an independent set $I$ of $n/4$ interior vertices (this is directly implied by the four color theorem \cite{RSST97}).
We split the edge removal process into two consecutive steps --- in the first step we iterate over the vertices of $I$ and remove a single edge adjacent to each vertex, and in the second step we iterate over the rest of the vertices in the same manner. Notice that after completing the first step, the sum of the degrees of the vertices in the second step is reduced by $n/4$, regardless of which specific edges were chosen. Thus, the sum of the degrees is reduced to $5.75N$, which implies the asserted bound.}
\end{theProof}
\begin{theorem}
Any triangulation embedded over a set $S$ of $N$ points in the plane contains $O^*(6.54^N)$ pseudo-triangulations.
\end{theorem}
\begin{theProof}{\!\!}
Let $\ptw{i}{T}$ denote the number of pseudo-triangulations with exactly $i$ pointed interior vertices that are contained in $T$.
We wish to bound $\sum_{i=0}^{n}\ptw{i}{T} = O^*\left(\max_i\left(\ptw{i}{T}\right)\right)$. Since $T$ has $2N+n-3$ edges and a pseudo-triangulation with $i$ pointed interior vertices has $2N+n-3-i$ edges (e.g., see \cite[Theorem 2.5]{RSS06}), we get the bound
\begin{align}
\ptw{i}{T} &\le \binom{2N+n-3}{2N+n-3-i} = O^*\left(\binom{3N}{i} \right) \nonumber \\[1mm]
&= O^*\left( \left(\frac{27}{a^a \cdot (3-a)^{3-a}} \right)^N\right)=: f_1(a), \label{eq:trivial}
\end{align}
where $a = i/N$ (for the last transition we have used Stirling's approximation). While this trivial method implies a good bound when $i$ is small, it yields a bound of $O^*\left(6.75^N\right)$ when $i \approx N$.
When $i$ is large, we can obtain a better bound by relying on Theorem \ref{th:PPTinTR}.
That is, we have $O(5.45^N)$ ways of choosing a unique edge for every vertex. For each such choice of edges, we have less than $\binom{N}{i}$ ways of choosing the subset of the edges that will actually be removed. Thus, we have
\begin{equation} \label{eq:nontrivial}
\ptw{i}{T} < O(5.45^N) \cdot \binom{N}{i}  = O^*\left(\left( \frac{5.45}{a^a \cdot (1-a)^{1-a}} \right)^N \right)=: f_2(a),
\end{equation}
where $a = i/N$.

\ignore{When $i$ is large, we can obtain a better bound by using Lemma \ref{le:bijection}, as follows.
There are $\binom{n}{i} < \binom{N}{i}$ ways to choose the set $P$ of the pointed interior vertices of the pseudo-triangulation. Similarly to the proof of Theorem \ref{th:PPTinTR}, we denote the degrees in $T$ of the vertices of $P$ as $d_1, \ldots, d_i$. As before, the number of pseudo-triangulations whose set of pointed interior vertices is exactly $P$ is at most $\prod_{j=1}^{i}d_j$. Since each of the $n-i$ vertices of $S'\setminus P$ (recall that $S'$ is the set of interior vertices) has a degree of at least 3 in $T$, we have $\sum_{j=1}^{i}d_j \le 2N+4n-6-3(n-i) < 3N+3i$ (there are $N+2n$ edges in $T$ that are not part of the convex hull). By also using the independent set trick from the proof of Theorem \ref{th:PPTinTR}, and noting that $P$ contains an independent set of size $i/4$, we obtain the bound
\begin{equation} \label{eq:nontrivial}
\ptw{i}{T} < \binom{N}{i} \cdot \left( \frac{3N+2.75i}{i}\right)^{i}  = O^*\left(\left( \frac{(3+2.75a)^a}{a^{2a} \cdot (1-a)^{1-a}} \right)^N \right), \mbox{ where $a = i/N$.}
\end{equation}
}

\begin{figure}[h]
\centering
\includegraphics[width=0.5\textwidth]{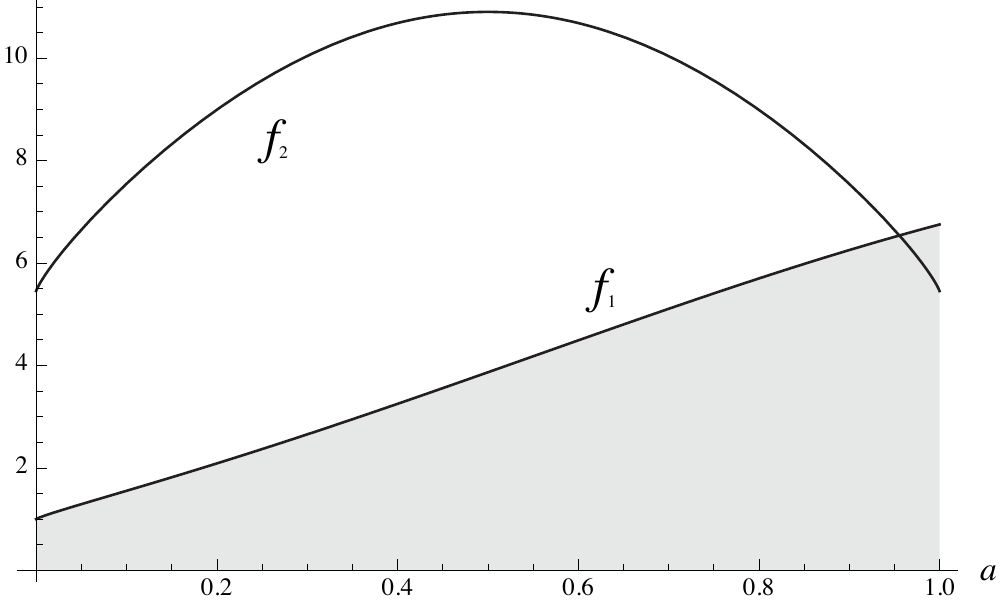}
\caption{\small \sf Bounds \eqref{eq:trivial} and \eqref{eq:nontrivial} coincide when $a\approx 0.955$.}
\label{fi:bounds}
\end{figure}

When $i$ is small we use bound \eqref{eq:trivial}, and when $i$ is large we use bound \eqref{eq:nontrivial}. The way these bounds behave is depicted in Figure \ref{fi:bounds}. The bounds coincide when $a\approx 0.955$, which implies $\ptw{i}{T} = O^*(6.54^N)$.
%This does not matter ! \footnote{The plot and the analysis were obtained using Wolfram$\mid$Alpha --- \url{www.wolframalpha.com}.}
\end{theProof}

\section{An upper bound for outdegree-1 orientations in planar graphs} \label{sec:Orient}

We give now the missing technical details in the proof of Theorem \ref{th:PPTinTR}.
Specifically, given a triangulation $T$ of $N$ points in the plane, we prove that there are $O(5.45^N)$ ways to choose a unique edge for every interior vertex of the triangulation, such that every edge is adjacent to its corresponding vertex.

Let $G$ be a planar graph with $N$ vertices. By a standard use of Euler's formula with a double counting argument, the sum of the vertex degrees of $G$ is less than $6N$. We call a partial orientation of the edges of $G$ an \emph{outdegree-1 orientation}, if every vertex in $G$ is incident to exactly one outgoing edge. To obtain an upper bound for the number of all outdegree-1 orientations of $G$, we count the number of possibilities for picking one outgoing edge from every vertex.  Let $d_i$ denote the degree of the vertex $v_i$. By using the arithmetic-geometric mean inequality, we have
\[\prod_{i=1}^{N} d_i \le \left(\frac{\sum_{i=1}^{N}d_i}{N}\right)^N < 6^N.\]
This number exceeds the number of outdegree-1 orientations, since some edges might be chosen as ``outgoing'' from both of their endpoints - in this case we say, that the edge forms a 2-cycle in the oriented graph. To overcome this inefficiency, we analyze the probability $P_\text{nc}$ that a random selection (of one outgoing edge for each vertex) contains no 2-cycles.
Notice that the number of valid partial orientations without 2-cycles is given by  $\left( \prod_{i=1}^{N} d_i \right) \cdot P_\text{nc}$.

\begin{figure}[h]
\centering
\includegraphics[width=0.48\textwidth]{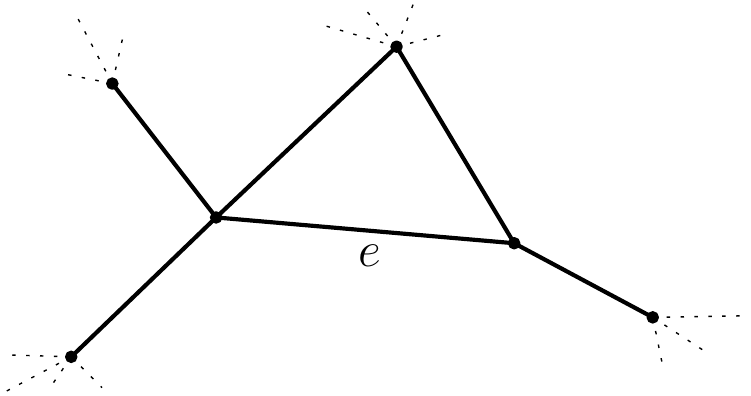}
\caption{\small \sf  A 2-extension based on the edge $e$ and with the signature $(3,4,(6,4),(5,4,6))$.}
\label{fi:2extensions}
\end{figure}

The framework of Buchin and Schulz~\cite{BS10} allows us to bound the probability $P_\text{nc}$. We give some intuition behind this approach, but for more technical details we redirect the reader to the original paper. Let $E_e$ be the event that a 2-cycle occurs in some random outdegree-1 orientation at edge $e$. If two edges $e$ and $e'$ share a vertex, then the events $E_e$ and $E_{e'}$ are dependent and \emph{mutually exclusive} (that is, at most one of the two events can occur in the same \mbox{outdegree-1} orientation). Based on these observations, we can derive a bound for $P_\text{nc}$ with an expression that depends on the distribution of the vertex degrees in the underlying graph~\cite[Lemma 1]{BS10}. More precisely, we consider subgraphs of $T$ that are composed of an edge $e=(u,v)$ and of all of the edges that have a common vertex with $e$. We refer to such a subgraph as a \emph{2-extension}, and say that the \emph{signature} of a 2-extension $X$ is the set of the degrees of the vertices that participate in $X$.
We represent a signature as a 4-tuple $(i, j, A, B)$, where $i$ and $j$ are the degrees of $u$ and $v$, $A$ is the set of degrees of the neighbors of $u$, and $B$ is the set of degrees of the neighbors of $v$ (there might be vertices that correspond to elements in both sets).
An example of a 2-extension is depicted in Figure \ref{fi:2extensions}.
Given a signature $s$ that appears $x_s$ times in $T$ (that is, $x_s$ edges of $T$ have $s$ as their signature), we set $f_s = x_s/N$.
The probability $P_\text{nc}$ can be expressed in terms of the $f_s$ variables, and
we thus have a linear maximization problem whose variables are the $f_s$.
The program contains three constraints on the signature distribution, which are obtained from simple double counting arguments.
Solving this program is problematic since there is an infinite number of possible signatures, and thus an infinite number of variables. Instead of solving the problem directly, we consider the corresponding dual program (a linear program with three variables and infinitely many constraints), whose solution gives $\sum_{i=1}^N \log d_i + \log P_\text{nc}$. Denoting the three variables of the dual program as $\lambda_1, \lambda_2,\lambda_3$, we obtain the objective function
\[\text{\bf Minimize } \lambda_1+\lambda_2+ 3\lambda_3.\]
For every signature $(i,j,A,B)$ with integers $3\le i \le j$, a set $A$ of $i-1$ integers $\ge 3$, and a set $B$ of $j-1$ integers $\ge 3$, we have the constraint
  \begin{multline}\label{eq:dual}
\log P_{ij}(A,B) + \frac{1}{4}  \left(\frac{\log i}{i}+\frac{\log j}{j}\right) + \frac{3}{4}  \left( \sum_{a_r \in A} \frac{\log a_r}{a_r(i-1)} +\sum_{b_r \in B} \frac{\log b_r}{b_r(j-1)}\right)\\ -  \lambda_1 \left(\frac{1}{i}+\frac{1}{j}\right) - \lambda_2   \left( \sum_{a_r \in A} \frac{1}{a_r(i-1)}+\sum_{b_r \in B} \frac{1}{b_r(j-1)}\right) -\lambda_3 \leq 0,
 \end{multline}
 where
 \[P_{ij}(A,B) :=  1 - \frac{1} {ij   \sqrt{ \prod\limits_{x\in A} \left( 1 - \frac{1}{i\cdot x} \right) \prod\limits_{y\in B} \left( 1 - \frac{1}{j\cdot y} \right) }  }.\]
%Exponentiating the LP-solution gives the desired upper bound.
An additional constraint is $\lambda_3\geq 0$.
For the full details on how to derive this LP formulation, see~~\cite{BS10}.

The main difficulty in solving the above linear program is the fact that there are infinitely many constraints. We present a solution to the program that is not necessarily optimal, but is feasible and hence bounds the optimal solution. To determine such a feasible point we restrict the dual program to finitely many constraints. Solving this derived LP yields the following solution $(S)$:
\begin{equation}\label{eq:candidate}
  \begin{aligned}
\lambda _1&=0, \\
\lambda _2&=.50906817, \\
\lambda _3&=.39507190.
\end{aligned}
\end{equation}

\begin{theorem}\label{thm:out1}
Any triangulation embedded over a set of $N$ points in the plane contains $O(5{.}45^N)$ outdegree-1 orientations.
\end{theorem}
\begin{theProof}{\!\!}
In Lemma \ref{lem:candidate} we prove that the solution $(S)$ is feasible. The value of $(S)$ is $\lambda_1+\lambda_2+ 3\lambda_3 = 0.50906817 + 3\cdot 0.39507190=1.69428387$. Thus, the value of the optimal solution is at most 1.69428387, implying
$\prod_{i=1}^{N} d_i \cdot P_\text{nc} = O(e^{1.69428387N})=O(5{.}45^N)$.
As stated above, the expression on the left-hand side is the maximum number of outdegree-1 orientations of any triangulation embedded over $N$ points. The proof does not use the planarity of the graph, but only the bound on the maximum number of edges.
\end{theProof}

\begin{lemma}\label{lem:candidate}
The candidate solution $(S)$ specified in \eqref{eq:candidate} is a feasible solution of the above linear program.
\end{lemma}
\begin{theProof}{\!\!}
The proof relies partially on numerical computations and brute force tests that were performed with computer algebra software. All computations (Mathematica script and pdf-file) can be downloaded at \url{http://cs.uni-muenster.de/u/schulz/outdegree1.zip}.

We have infinitely many constraints of the form \eqref{eq:dual} to check, since there is one such constraint for every signature tuple $(i,j,A,B)$. We now show that it suffices to consider only a finite subset of these constraints. Once we have a sufficiently small set of constraints, we use a computer program to verify that they are all satisfied by the values of $(S)$.

First, we show that if either $i$ or $j$ are large enough, then the corresponding constraint is automatically fulfilled, and can thus be removed.  The term $\log P_{ij}(A,B)$ is negative, so we may ignore it for now. Since $\left(3/4 \cdot \log a - \lambda_2 \right)/a$ is maximized for integers at $a=5$, we can assume as worst case, that the entries in $A,B$ are all $5$. Furthermore, since $1/x \cdot \log x$ is maximized for integers when $x=3$ (and recalling that $|A|=i-1$ and $|B|=j-1$), we notice that the expression
\begin{align}\label{eq:weakdual}
1/4  \cdot (\log(i)/i + \log(3)/3) + 3/2 \cdot \log(5)/5 -2/5 \cdot \lambda_2 - \lambda_3
\end{align}
is at least as large as the expression of the constraint. Since this expression is negative when $i\ge 38$, constraints that correspond to tuples with $i\ge38$ are necessarily satisfied. By symmetry, this is also the case when $j\ge 38$. Thus, it remains to check the satisfiability of constraints for which $i,j\le 37$.

The above proves that it suffices to consider a finite subset of $(i,j)$ pairs, but the size of this subset is still large.
We can reduce some of the remaining constraints by using the same analysis, but replacing $\log(3)/3$ in expression \eqref{eq:weakdual} with $\log(j)/j$. By examining when the modified expression is negative, we remain with the following set of $i,j$ pairs:

\begin{multline*}
\qquad\{\{3,3\},\{3,4\},\{4,4\},\{3,5\},\{4,5\},\{5,5\},\{3,6\},\{4,6\},\{5,6\},\{6,6\},\{3,7\},\\
\{4,7\},\{5,7\},\{6,7\},\{7,7\},\{3,8\},\{4,8\},\{5,8\},\{6,8\},\{7,8\},\{8,8\},\{3,9\},\{4,9\},\\
\{5,9\},\{6,9\},\{7,9\},\{8,9\},\{9,9\},\{3,10\},\{4,10\},\{5,10\},\{6,10\},\{7,10\},\{8,10\},\\
\{9,10\},\{3,11\},\{4,11\},\{5,11\},\{6,11\},\{7,11\},\{8,11\},\{3,12\},\{4,12\},\{5,12\},\\
\{6,12\},\{7,12\},\{8,12\},\{3,13\},\{4,13\},\{5,13\},\{6,13\},\{7,13\},\{3,14\},\{4,14\},\\
\{5,14\},\{6,14\},\{7,14\},\{3,15\},\{4,15\},\{5,15\},\{6,15\},\{3,16\},\{4,16\},\{5,16\},\\
\{6,16\},\{3,17\},\{4,17\},\{5,17\},\{6,17\},\{3,18\},\{4,18\},\{5,18\},\{3,19\},\{4,19\},\\
\{5,19\},\{3,20\},\{4,20\},\{5,20\},\{3,21\},\{4,21\},\{5,21\},\{3,22\},\{4,22\},\{3,23\},\\
\{4,23\},\{3,24\},\{4,24\},\{3,25\},\{4,25\},\{3,26\},\{4,26\},\{3,27\},\{4,27\},\{3,28\},\\
\{4,28\},\{3,29\},\{3,30\},\{3,31\},\{3,32\},\{3,33\},\{3,34\},\{3,35\},\{3,36\},\{3,37\}\}\qquad
\end{multline*}

To analyze the remaining cases, we reintroduce the term $\log P_{ij}(A,B)$ into \eqref{eq:dual}.
This means that we can no longer assume that the entries in $A,B$ are all $5$.
That is, even though we have a small set of possible values for $i$ and $j$, the number of constraints that we consider is once again infinite. To fix this, we show that it suffices to consider a finite set of possible values for the entries in $A,B$.

Consider an element $a\in A$, and notice that $\log P_{ij}(A,B)$ is increasing in $a$. Moreover, according to the above analysis, the sum of the other terms in \eqref{eq:dual} is increasing in $a$ in the range $[3,5]$.
Thus, we may assume that all the elements in $A$ and $B$ are at least $5$.

Next, we wish to find an upper bound for $a$. We reformulate the left-hand side of \eqref{eq:dual} as
\begin{align*}
I(a):=\log\left(1-\frac{1}{X \sqrt{1-1/(ia)}}\right)+\frac{3}{4} \frac{\log a}{ a(i-1)} -\lambda_2 \frac{1}{a(i-1)} +Y,
\end{align*}
where $X$ and $Y$ depend on $i,j,B$, and the other elements of $A$.
We wish to bound the integer $a_\text{max}$ that maximizes $I(a)$.
%NOT TRUE: Since $\frac{ d}{  da} I(a)$ is positive for $a=e$ and decreasing afterwards, $a_\ell$ will be the only root of $\frac{ d}{  da} I(a)$ larger than $e$.
We have
\begin{align*}
\dfrac{ d}{  da} I(a) = J_1(a) + J_2(a),
\end{align*}
where
\begin{align*}
J_1(a)&:=\left( 2 a (ai-1)\left(X\sqrt{1 - 1/(a i)} -1\right)  \right)^{-1},  \\
J_2(a) &:= \frac{\lambda_2+3/4(1-\log a)}{a^2(i-1)}.
\end{align*}
Notice that $X$ is minimized when all entries in $A,B$ are 3, and then its value is
\begin{align*}
X_\text{min} := ij(1 - 1/(3 i))^{(i - 2)/2}  (1 - 1/(3 j))^{(j - 1)/2}.
\end{align*}
%We observe that $\frac{ d}{  da} I(a)$ has a pole when $a=1/(X\sqrt{1 - 1/(a i)})$ and denote this value of $a$ as $a_p$. Moreover, the derivative $\frac{ d}{  da} I(a)$ is decreasing in the range $[a_p,\infty)$. Therefore, $I(a)$ obtains its maximum for some value $a_0 \in [a_p,\infty)$, where $a_0$ is a root of $\frac{ d}{  da} I(a)$.
%
Next, we notice that $J_1$ is positive and decreasing in $a$ (for every valid assignment of values to $i,j,X$). The behavior of $J_2$ for $i=5$ is depicted in Figure~\ref{fig:J2}. Changing the value of $i$ will only ``stretch" the graph of $J_2$, and thus, $J_2$: (i) has a root $a_0\approx5.3588$, (ii) is positive in the range $[e,a_0)$, and (iii) attains its only local minimum at $a_\text{dip}\approx8.83522$. We notice that $\dfrac{ d}{  da} I(a_\text{dip})$ is negative (for this, it suffices to check the extreme case where $X=X_\text{min}$ and all relevant pairs $i,j$), and thus, $a_\text{max}$ is the root of $J_1(a)+J_2(a)$ left of $a_\text{dip}$. Using computer algebra software we check for all remaining pairs $i,j$  whether there is only one root for $\dfrac{ d}{  da} I(a_\text{dip})$ when $X=X_\text{min}$, and then we determine as bound for $a_\text{max}$ as the smallest integer larger than this root. We find that we may assume that no element of $A,B$ is larger than 6 (i.e., that $a_\text{max}<6$).

\begin{figure}[h]
\centering
\includegraphics[width=0.73\textwidth]{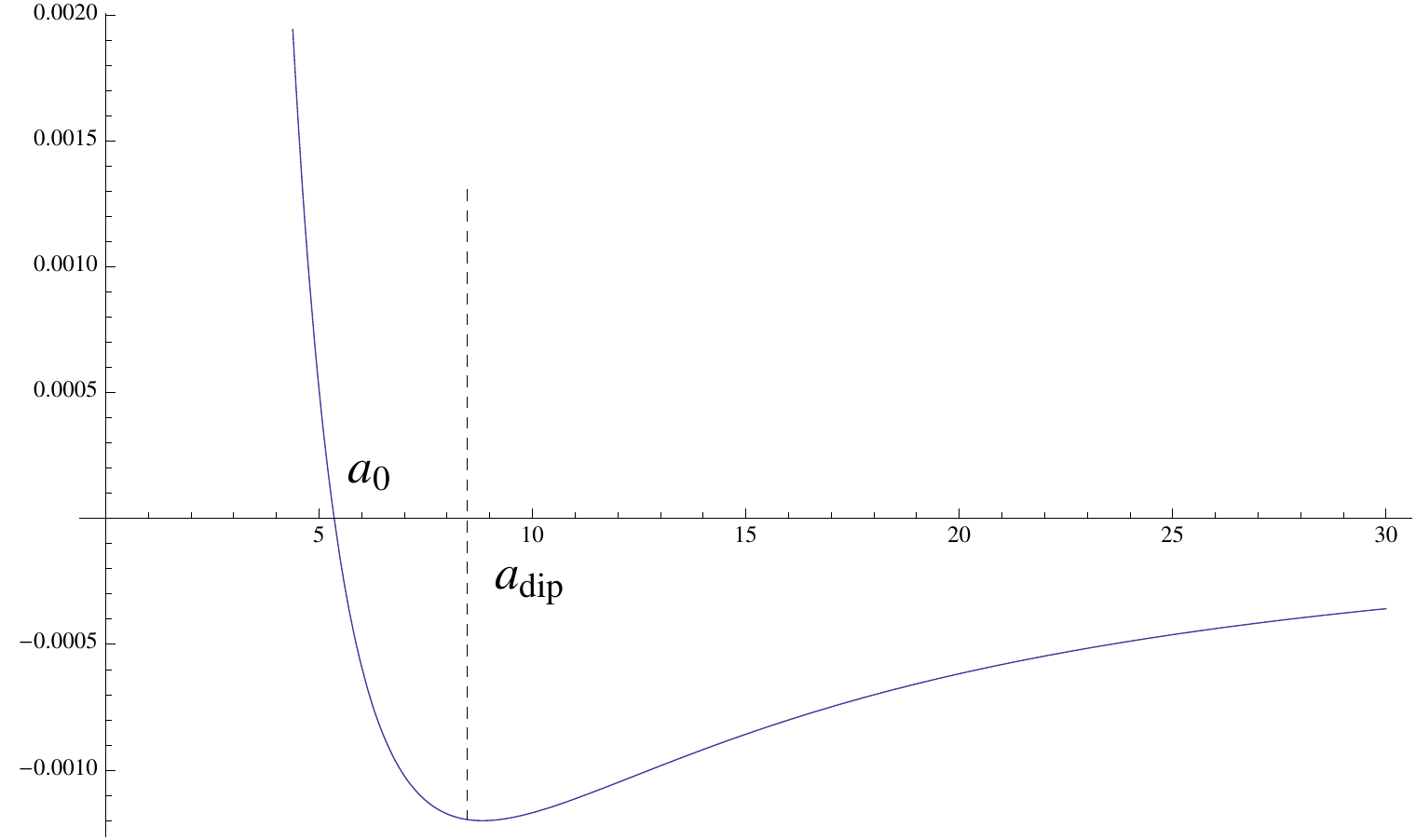}
\caption{\small \sf The graph of the function $J_2(a)$ for $i=5$.}
\label{fig:J2}
\end{figure}

Summing up, it suffices to check the remaining $i,j$ pairs with all possible combinations of $A,B$. This is possible since we may assume that $A$ and $B$ only contain elements from $\{5,6\}$. Notice that the order inside the sets $A,B$ is irrelevant. Our computations show that all of the constraints that correspond to the remaining tuples are satisfied by the above candidate solution.
\end{theProof}

\section{The number of pseudo-triangulations of a point set} \label{sec:conj}

There are various conjectures regarding the relations between $\ppt{S},\pt{S}$, and $\tri{S}$.
It is conjectured that $\ppt{S} \ge \tri{S}$ holds for any points set $S$~\cite{AOSS08,RRSS01}.
Similarly, Aichholzer~et al.~\cite{AOSS08} conjecture that for any set $S$,  subset of interior points $W \subset S$, and point $v \in W$, it holds that $\ptw{W}{S} \ge \ptw{W\setminus \{v\}}{S}$ (notice that this conjecture immediately implies $\ppt{S} \ge \tri{S}$).
Both of these variants have been verified for all sets of at most 10 points \cite{AK01}, for some interesting configurations of $N$ points \cite{AOSS08}, and various other insights have been observed regarding them. However, in the decade that has passed since the establishment of Theorem \ref{le:old}, no actual progress has been made in the from of improved bounds (and Theorem \ref{le:old} bounds the opposite direction).

In this section, we present a new observation related to Theorem \ref{le:old} (and to the above conjectures), and then rely on it to show that the bound $\ppt{S} \le 3^N \cdot \tri{S}$ is indeed not asymptotically tight.

\begin{figure}[h]
\centering
\includegraphics[width=0.7\textwidth]{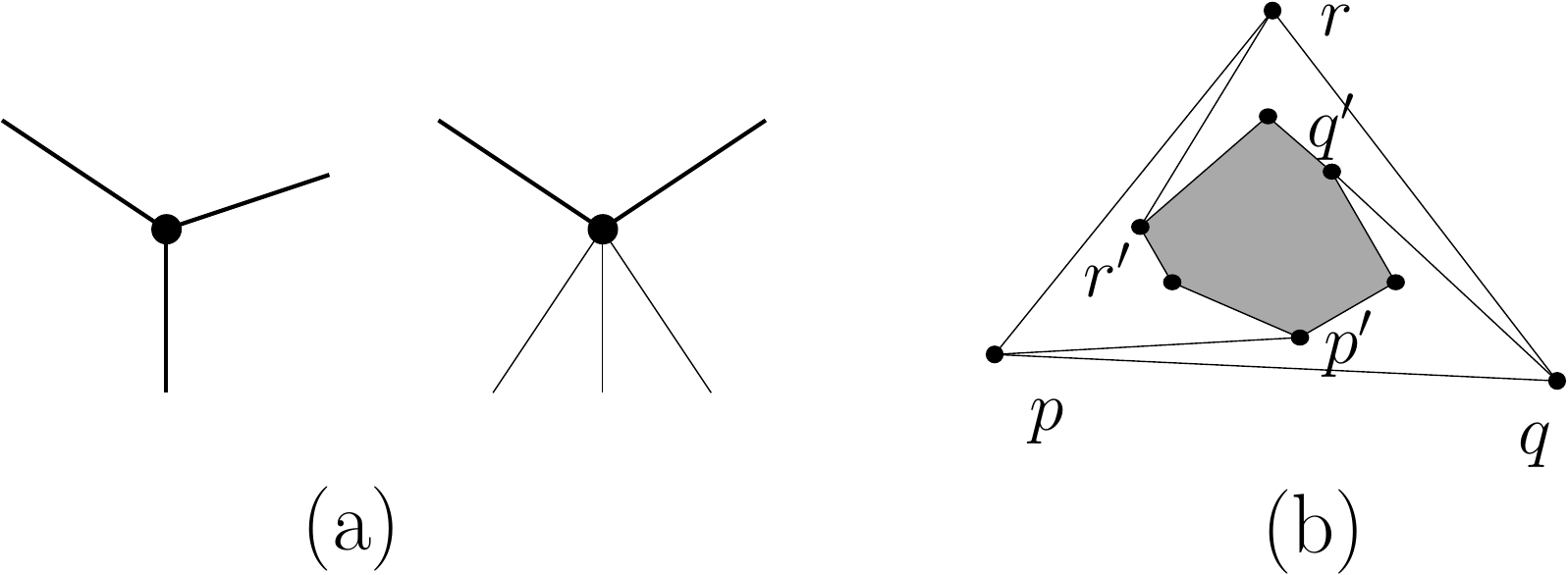}
%\begin{tabular}{cp{1cm}c}
%\includegraphics[width=0.25\textwidth]{deg34} &&
%\includegraphics[width=0.23\textwidth]{trianglehull} \\
%(a) && (b)
%\end{tabular}

\caption{\small \sf (a) Removing any edge incident to a vertex of degree 3 will make it pointed. For vertices of larger degrees there are at most two edges with this property.
(b) By inserting $p,q,r$ far enough from the convex hull of $S$ (the shaded part), we obtain a triangular convex hull.}
\label{fi:ppt1}
\end{figure}

\begin{observation} \label{ob:first}
For every set $S$ of $N$ points in the plane, there exists a point $v \in S'$ such that $\ptw{\{v\}}{S} < (8/3 + h/(3n)) \cdot \ptw{\emptyset}{S}$.
\end{observation}
\begin{theProof}{\!\!}
Returning to the second observation from the proof-sketch of Theorem \ref{le:old}, we notice that for a vertex $v$ to have three edges whose removal will make $v$ pointed, $v$ must have a degree of 3. If $v$ has a larger degree, there are at most two such edges (an example is depicted in Figure \ref{fi:ppt1}(a)). For some more intuition about this subject, see a discussion about \emph{separable edges}~\cite{SSW10}.

It is well known that any triangulation embedded over a set of $N$ points has fewer than $(2n+h)/3$ interior vertices of degree 3~\cite[Section 2]{SW06b}.
Therefore, every triangulation of $S$ has more than $(n-h)/3$ interior vertices of a larger degree. By the pigeonhole principle, there exists a vertex $v \in S'$ that has a degree larger than 3 in more than $(n-h)/(3n) \cdot \tri{S}$ triangulations of $S$. Combining this with the first observation from Theorem \ref{le:old} (for every pseudo-triangulation with $W$ as the set of its pointed interior vertices, we can insert a (unique) single edge to obtain a pseudo-triangulation
with $W \setminus \{v\}$ as the set of its pointed interior vertices), we obtain
\[\ptw{\{v\}}{S} < 2\cdot (n-h)/(3n) \cdot \tri{S} + 3\cdot (2n+h)/(3n) \cdot \tri{S} = (8/3 + h/(3n)) \cdot \ptw{\emptyset}{S}.\]
\end{theProof}

In the proof of the following Theorem \ref{th:ppt}, we show that this observation can be extended to cases where $W$ has a larger size. Moreover, at the end of this section we show how Observation \ref{ob:first} can be further improved. But first, since the bound in the observation gets better as the convex hull gets smaller, we derive the following lemma.
\begin{lemma} \label{le:quad}
Let $c>1$ be a constant such that every set $S$ with a triangular convex hull satisfies $\ppt{S} = O(c^{|S|})$.
Then $\ppt{S} = O(c^{|S|})$ also holds for sets $S$ with any size of a convex hull.
\end{lemma}
\begin{theProof}{\!\!}
Consider a set $S$ of $N$ points with no restriction on the size of its convex hull.
We pick three new points $p,q,r$, such that the convex hull of $p,q,r$ contains the convex hull of $S$ in its interior. We set $S^*=S \cup \{p,q,r\}$. Let $p'$ be the first point of $S$ that is hit when rotating  the segment $pq$ counterclockwise around $p$. The points $q'$ and $r'$ are defined symmetrically (see Figure~\ref{fi:ppt1}(b)).

Let $T$ be a pointed pseudo-triangulation of $S$. By inserting the edges $pq,pr,qr,pp',qq',rr'$ into $T$ (together with the points $p,q,r$) we obtain a pointed pseudo-triangulation of $S^*$ that contains $T$. This implies that we can map every pointed pseudo-triangulation of $S$ to
a distinct pointed pseudo-triangulation of $S^*$, and thus, $\ppt{S} \le \ppt{S^*}$.
The lemma then follows since $\ppt{S^*} = O(c^{N+3}) = O(c^N)$.
 \end{theProof}

\begin{theorem} \label{th:ppt}
For any set $S$ of $N$ points in the plane $\ppt{S} = O^*(2.97^N) \cdot \tri{S}$.
\end{theorem}
 \begin{theProof}{\!\!}
 By Lemma \ref{le:quad}, it suffices to consider point sets with a triangular convex hull.
 Thus, by Observation \ref{ob:first}, there exists a point $v_1 \in S'$ such that $\ptw{\{v_1\}}{S} < (8/3 + 1/n) \cdot \ptw{\emptyset}{S}$.

 Next, notice that we can use the trick from the proof of Observation \ref{ob:first} once again. Recall that every triangulation has at least $(n-h)/3$ interior vertices of degree larger than 3. Removing a single edge can make one vertex pointed and reduce the degree of another vertex down to 3. Thus, every pseudo-triangulation that is counted in $\ptw{\{v_1\}}{S}$ contains at least $(n-h)/3-2$ interior non-pointed vertices of degree larger than 3. The same analysis implies the existence of an interior vertex $v_2$ such that
\begin{align*}
\ptw{\{v_1,v_2\}}{S} &< \left(3\cdot \frac{(2n+h)/3+1}{n-1} + 2\cdot \frac{(n-h)/3-2}{n-1}\right) \cdot \ptw{\{v_1\}}{S} \\
&= \frac{8n/3}{n-1}\cdot \ptw{\{v_1\}}{S}.
\end{align*}
Repeating this iteratively, we get in the $(i+1)$-th step at least $(n-h)/3-2i$ interior non-pointed vertices of degree larger than 3, which implies
\[\ptw{\{v_1,\ldots,v_{i+1}\}}{S} < \frac{8n/3-i+1}{n-i}\cdot \ptw{\{v_1,\ldots,v_i\}}{S}.\]
We can perform $n/6$ steps of this process and then use Theorem \ref{le:old}, to obtain
\begin{equation}\label{eq:withProd}
\ppt{S} < \left(\prod_{i=0}^{n/6}\frac{8n/3-i+1}{n-i}\right) \cdot 3^{5n/6} \cdot \tri{S}.
\end{equation}
Next, we notice that
\[ \prod_{i=0}^{n/6}\frac{8n/3-i+1}{n-i} = \frac{(5n/6)!}{n!}\cdot \frac{(8n/3+1)!}{(5n/2+1)!}. \]
By using Stirling's approximation (that is, $m! = O^*\left((m/e)^m\right)$), we have
\begin{equation} \label{eq:prod}
\prod_{i=0}^{n/6}\frac{8n/3-i+1}{n-i} = O^*\left(\left( \frac{(5/6)^{5/6}\cdot(8/3)^{8/3}}{(5/2)^{5/2}}\right)^N\right).
\end{equation}
Finally, by combining \eqref{eq:withProd} and \eqref{eq:prod}, we obtain
\[\ppt{S} = O^*\left( \left( \frac{(5/6)^{5/6}\cdot(8/3)^{8/3} \cdot 3^{5/6}}{(5/2)^{5/2}}\right)^N\right) \cdot \tri{S} = O^*(2.97^N) \cdot \tri{S}. \]
\end{theProof}

Combining Theorem \ref{th:ppt} with the bound $\tri{N} < 30^N$ from \cite{SS10} results in the following corollary
\begin{corollary}
$\ppt{N} =O^*(89.1^N)$.
\end{corollary}

\noindent{\bf Remark.} We believe that this technique can yield more significant improvements, since our analysis is not tight in several places: (i) There exist vertices with no separable edge adjacent to them. Such vertices get a multiplier of 0 instead of 2. (ii) When removing a vertex of degree 3 (which is the common case in our analysis), we remove at most one vertex of a larger degree, instead of two, which should provide a better ratio for the following round. (iii) We will now show that Observation \ref{ob:first} is far from being tight, though our proof does not easily extend to the following rounds (i.e., when there is more than a single pointed interior vertex).

We conclude this section by presenting an improvement for Observation \ref{ob:first}.
\begin{observation} \label{ob:second}
For every set $S$ of $N$ points in the plane, there exists a point $v \in S'$ such that $\ptw{\{v\}}{S} \le (12/5 + h/(10n)) \cdot \ptw{\emptyset}{S}$.
\end{observation}
\begin{theProof}{\!\!}
In the proof of Observation \ref{ob:first} we relied on the worst case bound on the number of vertices of degree 3 in a triangulation.
However, it can be shown that the average case is much better. For any point set $S$, on average, a triangulation of $S$ has at most $(2n+h/2)/5$ vertices of degree 3~\cite{SSW10} . That is, a uniformly chosen triangulation (from the set of all triangulations of $S$) is expected to have at most $(2n+h/2)/5$ vertices of degree 3. Using the same analysis as in the proof of \ref{ob:first}, we notice that there exists some $v\in S'$, such that
\[\ptw{\{v\}}{S} \le 2\cdot \frac{3n-h/2}{5n} \cdot \tri{S} + 3\cdot \frac{2n+h/2}{5n} \cdot \tri{S} = \left(\frac{12}{5} + \frac{h}{10n}\right) \cdot \ptw{\emptyset}{S}.\]
\end{theProof}

\noindent{\bf Acknowledgements.}
We wish to thank Micha Sharir for his insightful remarks and for making this work possible.
We also wish to thank Oswin Aichholzer for providing helpful information about the previous results in this topic (he also noticed the bound $\pt{N} < 120^N$, independently from us).
Finally, we thank the anonymous referees for the many helpful remarks.

%%%%%%%%%%%%%%%%%%%%%%%%%%%%%%%%%%%%%%%%%%%%%%%%%%%%%%%%%%%%%%%%%%%%%%%%%%%%%%%%%

%\ignore{
%%%%%%%%%%%%%%%%%%%%%%%%%%%%%%%%%%%%%%%%%%%%%%Misceallaneous Material
%\input{app.tex}
%}

\end{document}